\documentclass[aps,prb,twocolumn,preprintnumbers,amsmath,amssymb,superscriptaddress]{revtex4}%

\usepackage{graphicx}%
\usepackage{dcolumn}
\usepackage{amsmath}
\usepackage{color}

\begin{document}

\title{Pressure weakens coupling strength in In and Sn elemental superconductors}

\author{Rustem~Khasanov}
 \email{rustem.khasanov@psi.ch}
 \affiliation{Laboratory for Muon Spin Spectroscopy, Paul Scherrer Institut, CH-5232 Villigen PSI, Switzerland}

\author{Giovanni A. Ummarino}
\email{giovanni.ummarino@polito.it}
 \affiliation{Istituto di Ingegneria e Fisica dei Materiali, Dipartimento di Scienza Applicata e Tecnologia, Politecnico di Torino, Corso Duca degli Abruzzi 24, 10129 Torino, Italy}
 \affiliation{National Research Nuclear University MEPhI (Moscow Engineering Physics Institute), Kashira Hwy 31, Moscow 115409, Russia}

\begin{abstract}
Pressure dependence of  the thermodynamic critical field $B_{\rm c}$ in elemental indium (In) and tin (Sn) superconductors was studied by means of the muon-spin rotation/relaxation. Pressure enhances the deviation of $B_{\rm c}(T)$ from the parabolic behavior, expected for a typical type-I superconductor, suggesting a weakening of the coupling strengths $\alpha=\langle\Delta\rangle /k_{\rm B}T_{\rm c}$ ($\langle\Delta\rangle$ is the average value of the superconducting energy gap, $T_{\rm c}$ is the transition temperature and $k_{\rm B}$ is the Boltzmann constant). As pressure increases from 0.0 to $\simeq 3.0$~GPa, $\alpha$ decreases linearly approaching the limiting weak-coupling BCS value $\alpha_{\rm BCS}=1.764$. Analysis of the data within the framework of the Eliashberg theory reveals that only part of the pressure effect on $\alpha$ can be attributed to the effect of hardening of the phonon spectra, which is reflected by a decrease of the electron-phonon coupling constant. Nearly 40\% of the effect is caused by increased anisotropy of the superconducting energy gap.
\end{abstract}

\maketitle


Pressure is known to be one of the important parameters in superconductor physics.  It allows the discovery of many new materials that do not superconduct or may not even exist under ambient pressure conditions, but exhibit significantly high critical temperatures under applied pressure.\cite{Lorenz_2005, Schilling_book_2007, Schilling_JPCS_2008} In particular, pressure experiments enable the discovery of completely new classes of high-temperature superconducting materials, with transition temperatures exceeding the liquid nitrogen level, like nickelates,\cite{Sun_Nature_2023} or even approaching room temperature like hydrides.\cite{Somayazulu_PRL_2019, Drozdov_Nature_2019, Semenok_arxiv_2024} At the same time, pressure serves as a clean tuning parameter that, without causing structural changes, fine-tunes the crystal lattice and, as a consequence, allows tracking the corresponding changes in various superconducting properties of the material.

In most conventional phonon-mediated superconductors, the application of pressure leads to a decrease in both, the superconducting transition temperature $T_{\rm c}$ and the superconducting energy gap $\Delta$.\cite{Lorenz_2005, Schilling_book_2007, Schilling_JPCS_2008, Franck_PRL_1968, Svistunov_SovUspekhi_1987, Zhu_APL_2015, Du_PRL_2024} At first glance, it is expected that these two quantities would change proportionally to each other. This expectation is dictated, in particular, by the universal relation established within BCS theory:\cite{BCS_PR_1957, Tinkham_book_1975}
\begin{equation}
\alpha_{\rm BCS}=\Delta/k_{\rm B}T_{\rm c}=e^{\gamma_E}/\pi\simeq 1.764.
 \label{eq:alpha_BCS}
\end{equation}
Here, $\gamma_{\rm E}$ and $k_{\rm B}$ are the Euler and Boltzmann constants, respectively. Experimentally, however, it was found that $\Delta$ decreases faster than $T_{\rm c}$, suggesting that $\alpha=\Delta/k_{\rm B}T_{\rm c}$ is, in fact, pressure-dependent.\cite{Franck_PRL_1968, Svistunov_SovUspekhi_1987, Galkin_PSS_1969, Nedellec_SSComm_1973}

Note, that in superconductor physics $\alpha=\Delta/k_{\rm B}T_{\rm c}$ has a special meaning and is offen called the `coupling strength'. By comparing $\alpha$s with the universal BCS value $\alpha_{\rm BCS}$, the superconductors are divided into strong-coupling ($\alpha \gg 1.764$), intermediate-coupling ($\alpha \gtrsim 1.764$) and weak-coupling ($\alpha\simeq 1.764$) classes. The BCS theory implies that in a case of a single, uniform (in both real and momentum space) order parameter, $\alpha_{\rm BCS}\simeq1.764$ sets the lower bound for possible coupling strength values. In other words, weaker coupling {\it i.e.}, $\alpha < \alpha_{\rm BCS}$ becomes physically impossible.

Regarding the above-mentioned faster decrease of $\Delta$ compared to $T_{\rm c}$, two important consequences are expected to follow:
 (i) Pressure reduces the coupling strength $\alpha$ and moves the superconductor into the weak-coupling direction.
 (ii) As $\alpha$ approaches the weak-coupling BCS value $\alpha_{\rm BCS} = 1.764$, the coupling strength should saturate and remain unchanged with further increases in pressure.
The first statement was indeed confirmed in tunneling studies of various conventional superconductors, suggesting the universality of the trend. Experiments reveal that moderate pressures (up to $\sim 2.0$~GPa) lead to substantial decrease in $\alpha$  in various single-element and binary superconducting materials.\cite{Franck_PRL_1968, Svistunov_SovUspekhi_1987, Galkin_PSS_1969, Nedellec_SSComm_1973}
The second trend, namely the saturation of $\alpha$ as it approaches $\alpha_{\rm BCS}$, has not been experimentally confirmed so far. On the contrary, measurements of the thermodynamic critical field $B_{\rm c}$ in elemental aluminum reveal that $\alpha$ may decrease below the $\alpha_{\rm BCS}$ level due to enhanced anisotropy of the superconducting energy gap.\cite{Khasanov_Aluminum_PRB_2021}

From the theoretical side, Leavens and Carbotte, Ref.~\onlinecite{Leavens_CanJPhys_1972}, have shown that the effect of pressure on the energy gap in conventional phonon-mediated superconductors is expected to be two-fold. First, pressure decreases the mean gap value much more significantly than $T_{\rm c}$ due to the effect of phonon hardening. Second, pressure is expected to increase the gap anisotropy, {\it i.e.}, the ratio between the largest and smallest energy gap values. This suggests that in studies of the pressure effect on coupling strength $\alpha$, both of the aforementioned contributions need to be considered.

\begin{figure*}[htb]
\includegraphics[width=1.0\linewidth]{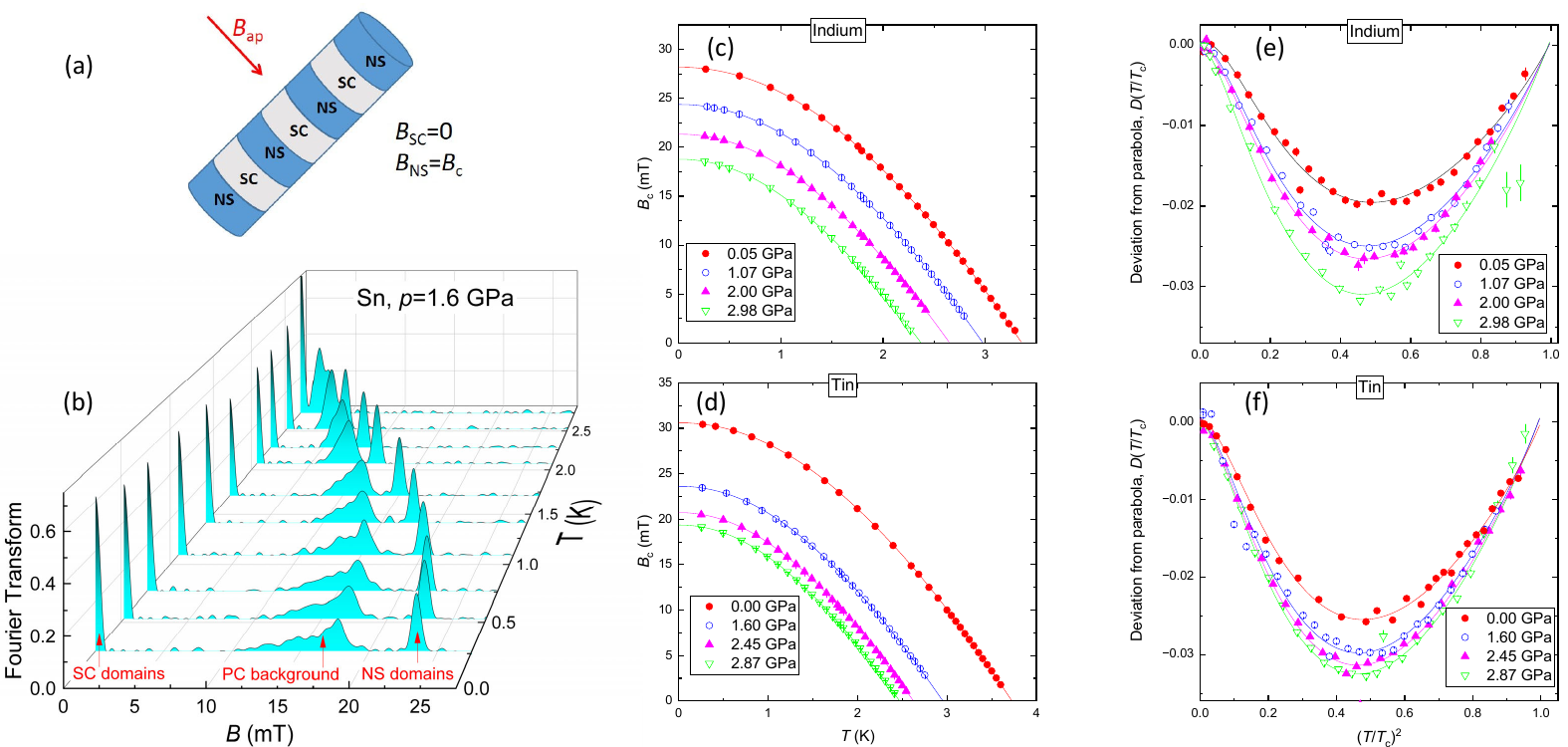}
%
\caption{(a) The schematic representation of separation of a type-I cylindrical sample into the normal state (NS) and the superconducting (SC) domains. The magnetic field in NS domains is equal to the thermodynamic critical field, $B_{\rm NS}=B_{\rm c}$. The field in SC domains is equal to zero, $B_{\rm SC}=0$. (b) Fourier transforms of TF-$\mu$SR data for the tin sample measured at $p=1.60$~GPa. Peaks at $B=0$, $B=B_{\rm c}$, and $B\simeq B_{\rm ap}$ correspond to the contributions of the superconducting domains, normal state domains, and the background caused by the pressure cell, respectively. (c) Temperature dependencies of the thermodynamic critical field $B_{\rm c}$ in elemental indium measured at pressures $p=0.05$, 1.07, 2.09, and 2.98~GPa.
(d) $B_{\rm c}(T)$ dependencies in elemental tin measured at $p=0.0$, 1.60, 2.45, and 2.87~GPa.  (e) Deviation functions $D({T/T_{\rm c}})=B_{\rm c}(T/T_{\rm c})-B_{\rm c}(0)\cdot[1-(T/T_{\rm c})^2]$ as obtained for the indium sample. (f) $D(T/T_{\rm c})$ dependencies for the tin sample.  The solid lines in panels (c)-(f) are fits of phenomenological $\alpha-$model to the $B_{\rm c}(T)$ data (see Ref.~\onlinecite{Supplementary_Information} for details). }
 \label{fig:Bc_Deviation}
\end{figure*}

In this paper, we studied the effect of pressure on the superconducting energy gap in elemental indium (In) and tin (Sn).
The average values of the superconducting energy gap ($\langle\Delta\rangle$) were determined from measurements of the temperature evolution of  the thermodynamic critical field $B_{\rm c}$ using the muon-spin rotation/relaxation technique. The analysis of the experimental data within the phenomenological $\alpha$-model of Padamsee {\it et al.}\cite{Padamsee_JLTP_1973, Johnston_SST_2013} suggests that an increase in pressure from $p=0.0$ to $\simeq 3.0$~GPa leads to a decrease in $\alpha=\langle\Delta\rangle/k_{\rm B}T_{\rm c}$ from 1.82 to 1.76 for indium and from 1.86 to 1.765 for tin, respectively. A simple model based on Eliashberg theory allows us to distinguish between two contributions, namely the enhancement of gap anisotropy and phonon hardening effects.


Sn and In samples were prepared from commercially available solid pieces ($\simeq 3-4$~mm in size and 99.999\% pure). Pieces of `soft' In and Sn metal were placed inside the 5~mm in diameter ($d=5$~mm) pressure cell channel and pressed with a force of $\sim1-1.5$~ton. Following this procedure, the metal fills the pressure cell channel and forms cylindrically shaped samples. The amount of material was chosen to achieve the 'compressed` sample height approximately 15~mm ($h\simeq 15$~mm). No pressure medium was used. The pressure cell consisted of three cylinders (three-wall pressure cell), where under ambient conditions each inner cylinder remains precompressed by the outer one. The construction and the characteristics of the three-wall pressure cell are described in Ref.~\onlinecite{Khasanov_Three-wall-cell_HPR_2022}.

The muon-spin rotation/relaxation ($\mu$SR) under pressure experiments were performed at the $\mu$E1 beamline using the General Purpose Decay (GPD) spectrometer, Paul Scherrer Institute, PSI Villigen, Switzerland.\cite{Khasanov_HPR_2016, Khasanov_JAP_2022} The $^4$He cryostat equipped with the $^3$He inset was used.  The external magnetic field $B_{\rm ap}$ was applied perpendicular to the initial muon-spin direction, corresponding to the transverse-field (TF-$\mu$SR) geometry. The experiments were conducted in the temperature range of 0.25 to 4.0~K and in the field range of 0.5 to 40 mT.

The TF-$\mu$SR measurements were performed in the intermediate state, {\it i.e.}, when the type-I superconducting sample is separated on the normal state (NS) and the superconducting (SC) [{\it i.e.} Meissner domains, see {\it e.g.}, Refs.~\onlinecite{Poole_Book_2014, deGennes_Book_1966, Kittel_Book_1996, Prozorov_PRL_2007, Prozorov_NatPhys_2008, Khasanov_Bi-II_PRB_2019, Karl_PRB_2019, Khasanov_Ga-II_PRB_2020, Khasanov_AuBe_PRR_2020} and schematic in Fig.~\ref{fig:Bc_Deviation}~(a)]. The magnetic field $B_{\rm ap}$ was applied perpendicular to the cylindrical axis of the sample. In this geometry the sample's demagnetization factor is estimated to be $n=(2+d/\sqrt{2} h)^{-1}\simeq 0.45$,\cite{Prozorov_PRAppl_2018} and the intermediate state is expected to form for applied fields in the range $B_{\rm c}<B_{\rm ap}\lesssim 0.55\cdot B_{\rm c}$.    The modified $B-T$ scan measuring scheme, as discussed in Refs.~\onlinecite{Karl_PRB_2019, Khasanov_AuBe_PRR_2020}, was used. At each particular temperature, the measured points were reached by first increasing $B_{\rm ap}$ above $B_{\rm c}$  ($B_{\rm ap}\simeq 35$~mT) and then decreasing it back to the measurement field.  The $B-T$ points were taken along the $\simeq 0.7\cdot B_{\rm c}(T)$, and $0.8\cdot B_{\rm c}(T)$ lines.
The TF-$\mu$SR data analysis procedure is described in Ref.~\onlinecite{Supplementary_Information}.



The magnetic field distribution in type-I superconductor in the intermediate state, which is probed directly by means of TF-$\mu$SR,  consists of two peaks corresponding to the response of the domains remaining in the superconducting Meissner state ($B=0$) and in the intermediate state ($B = B_{\rm c}>B_{\rm ap}$) [see the schematic representation at Fig.~\ref{fig:Bc_Deviation}~(a)]. Consequently, in TF-$\mu$SR experiments the value of $B_{\rm c}$ is directly and very precisely determined by measuring the position of $B>B_{\rm ap}$ peak.\cite{Khasanov_Bi-II_PRB_2019, Karl_PRB_2019, Khasanov_Ga-II_PRB_2020, Khasanov_AuBe_PRR_2020, Egorov_PRB_2001, Leng_PRB_2019, Gladisch_HypInteract_1979, Grebinnik_JETP_1980,  Beare_PRB_2019, Kozhevnikov_JSNM_2020}

The Fourier transforms of TF-$\mu$SR data for superconducting Sn sample measured at $p\simeq$~1.6~Gpa are shown in Fig.~\ref{fig:Bc_Deviation}~(b). The figure represents part of the experimental data accumulated at $B-T$ scan with $B_{\rm ap}\simeq0.7\cdot B_{\rm c}(T)$. Note that in addition to $B=0$ and $B=B_{\rm c}$ peaks corresponding to the response of the sample, the background peak caused by muons stopped in the pressure cell walls [denoted as 'PC background` in Fig.~\ref{fig:Bc_Deviation}~(b)] is also seen. The mean value of the background field is equal to $B_{\rm ap}$, while the broadening of the background signal is caused by the influence of the sample's stray field on the pressure cell walls.\cite{Maisurafze_PRB_2011}

\begin{figure}[htb]
\includegraphics[width=0.8\linewidth]{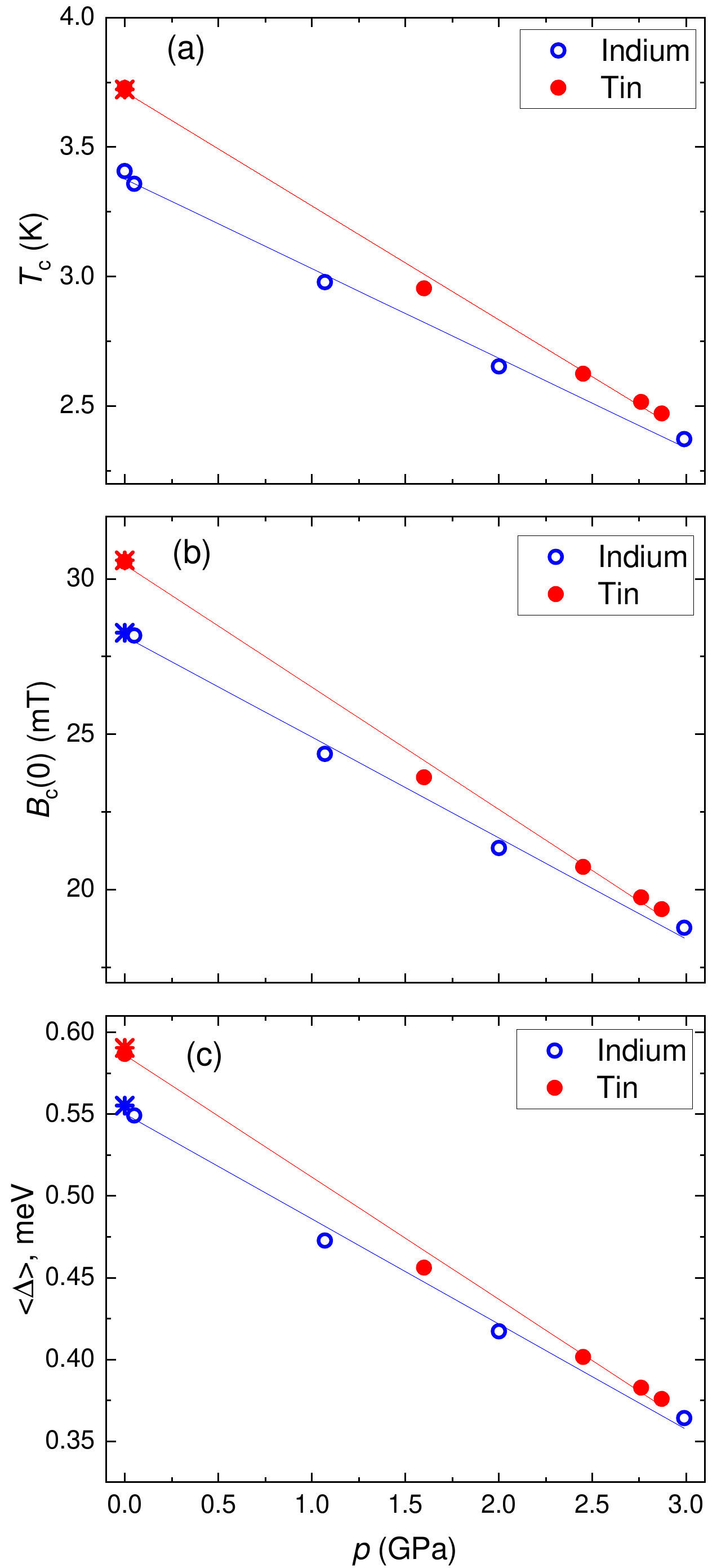}
%
\caption{Pressure dependence of: (a) the superconducting transition temperatures $T_{\rm c}$; (b) the zero-temperature values of the thermodynamic critical fields $B_{\rm c}(0)$; and (c) the mean values of the superconducting energy gaps $\langle\Delta\rangle$ for In and Sn samples. These quantities are obtained from the analysis of $B_{\rm c}(T,p)$ data within the framework of the phenomenological $\alpha-$model of Padamsee {\it et al.}\cite{Padamsee_JLTP_1973, Johnston_SST_2013} The circles correspond to the data obtained in the present study. The black asterisks represent parameters obtained from the analysis of the data of Finnemore and Mapother.\cite{Finnemore_PR_1965} The solid lines are linear fits. }
 \label{fig:Parameters_vs_p}
\end{figure}

The temperature dependencies of the thermodynamic critical field $B_{\rm c}$ at pressures ranging from $p=0.0$ to $\simeq3$~GPa are presented in Figs.~\ref{fig:Bc_Deviation}~(c) and (d) for the In and Sn samples, respectively.
Deviations of the $B_{\rm c}$ {\it vs.} $T$ curves from the parabolic function $D(T/T_{\rm c})=B_{\rm c}(T/T_{\rm c})/B_{\rm c}(0)-[1-(T/T_{\rm c})^2]$, where $B_{\rm c}(0)$ is the zero-temperature value of the thermodynamic critical field, are shown in Figs.~\ref{fig:Bc_Deviation}~(e) and (f).
The analysis of $B_{\rm c}(T,p)$ dependencies was performed within the framework of the phenomenological $\alpha$-model of Padamsee {\it et al.},\cite{Padamsee_JLTP_1973, Johnston_SST_2013} allowing the $B_{\rm c}(T)$ dependencies to be analyzed with only three independent parameters: $T_{\rm c}$, $B_{\rm c}(0)$, and $\Delta$ (see Ref.~\onlinecite{Supplementary_Information} for details). 
Fits of $\alpha-$model to the $B_{\rm c}(T)$ data are shown by solid lines in Figs.~\ref{fig:Bc_Deviation}~(c)-(f).

\begin{table*}[htb]
\caption{\label{tab1} Pressure dependencies of thermodynamic parameters for In and Sn samples. $T_{\rm c}(p=0)$ is the superconducting transition temperature at $p=0$,  $B_{\rm c}(0,p=0)$ is the zero-temperature zero-field value of the thermodynamic critical field, and $\langle\Delta(p=0)\rangle$ is the average zero-pressure value of the superconducting energy gap.}
\begin{tabular}{c|ccc|ccc|ccccc}
\hline
\hline
Sample&$T_{\rm c}(p=0)$ &${\rm d}T_{\rm c}/{\rm d} p$ & ${\rm d} \ln T_{\rm c}/{\rm d} p$&
       $B_{\rm c}(0,p=0)$&${\rm d}B_{\rm c}(0)/{\rm d}p$&${\rm d} \ln B_{\rm c}(0)/{\rm d}p$&
       $\langle\Delta(p=0)\rangle$&${\rm d}\langle\Delta\rangle /{\rm d}p$&${\rm d}\ln\langle\Delta\rangle /{\rm d}p$\\
      & (K) &(K/GPa)&(1/GPa)&
       (mT)&(mT/Gpa)&(1/GPa)&
       (mev)&(meV/GPa)&(1/GPa)  \\
\hline
Indium & 3.39(2)& -0.346(1)& -0.102(1)&28.15(5)&-3.25(6)& -0.115(2)&0.550(4)&-0.0643(5)&-0.117(2) \\
Tin    & 3.71(2)& -0.440(1)& -0.119(1)&30.45(6)&-3.94(5)& -0.129(2)&0.586(6)&-0.0748(5)&-0.126(2)\\
\hline
\end{tabular}
\end{table*}

Figure~\ref{fig:Parameters_vs_p} shows dependencies of $T_{\rm c}$, $B_{\rm c}(0)$, and the average value of the superconducting gap ($\langle\Delta \rangle$) on the applied pressure. Asterisks correspond to the values obtained from fits of $B_{\rm c}(T)$ curves for indium and tin reported by Finnemore and Mapother in Ref.~\onlinecite{Finnemore_PR_1965}. It should be noted that our experiments were performed on nonoriented metallic samples, so the value of the superconducting gap corresponds to a mean ({\it i.e.}, averaged) $\langle\Delta \rangle$ value.  Figure~\ref{fig:Parameters_vs_p} suggests, that for both In and Sn samples, all three thermodynamic quantities decrease nearly linearly with increasing pressure. The solid lines represent linear fits, with the parameters summarized in Table~\ref{tab1}.

\begin{figure}[htb]
\includegraphics[width=0.8\linewidth]{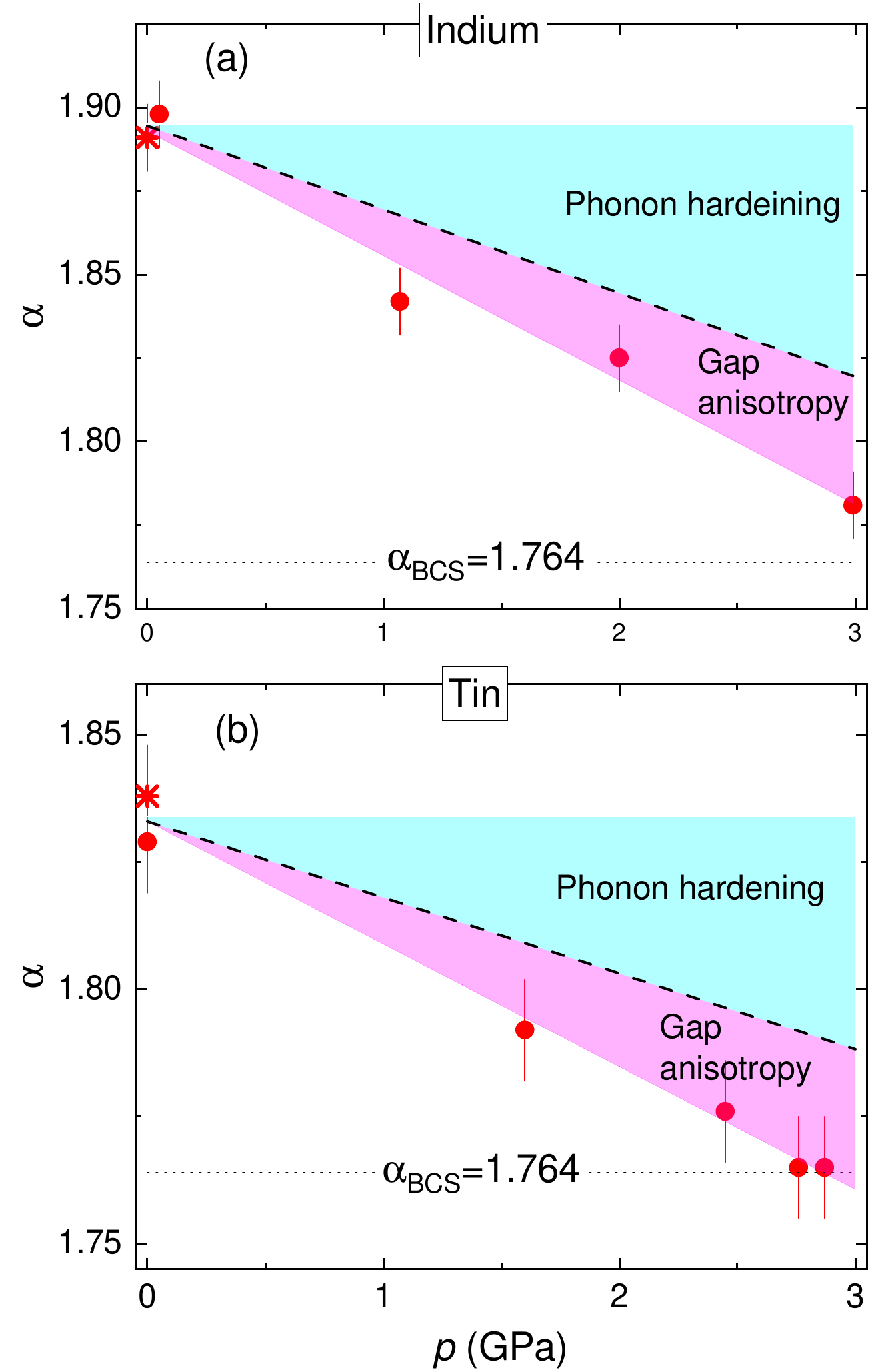}
%
\caption{(a) Dependence of the coupling strength $\alpha=\Delta/k_{\rm B}T_{\rm c}$ on applied pressure $p$ for the indium sample. Experimental data are represented by red circles. The 'phonon hardening' and 'anisotropic' contributions are shown in light blue and pink, respectively. (b) The same as in panel (a), but for the tin sample. The black asterisks represent parameters obtained from the analysis of the data of Finnemore and Mapother.\cite{Finnemore_PR_1965}}
 \label{fig:alpha_vs_p}
\end{figure}

From the data presented in Fig.~\ref{fig:Parameters_vs_p} and Table~\ref{tab1}, the following three important points emerge:\\
 (i) The zero-pressure values of the superconducting transition temperature $T_{\rm c}$, the thermodynamic critical field $B_{\rm c}(0)$, and the averaged superconducting energy gap $\langle\Delta\rangle$, as well as the pressure slopes of $T_{\rm c}$ and $\langle\Delta\rangle$, are in a good agreement with the literature data.\cite{Finnemore_PR_1965, Carbotte_RMP_1990, Poole_Book_2014, Lorenz_2005, Schilling_book_2007, Schilling_JPCS_2008, Franck_PRL_1968, Svistunov_SovUspekhi_1987, Galkin_PSS_1969, Nedellec_SSComm_1973}\\
 (ii) With increasing pressure, the superconducting energy gap $\langle\Delta\rangle$ decreases faster than the transition temperature $T_{\rm c}$, in agreement with the results reported in the literature.\cite{Svistunov_SovUspekhi_1987, Galkin_PSS_1969, Nedellec_SSComm_1973}  \\
 (iii) The relative pressure shifts of the thermodynamic critical field $B_{\rm c}(0)$ and the superconducting energy gap  $\langle\Delta\rangle$ are the same within experimental uncertainties. This could be due to the fact that $B_{\rm c}(0)$ is the measure of the energy, which has to be supplied to the material to destroy superconductivity. This implies that both $T_{\rm c}$ and $\langle\Delta\rangle$ are subject to similar energy scales in conventional phonon-mediated superconductors.


Figure~\ref{fig:alpha_vs_p} shows the dependencies of $\alpha=\langle\Delta\rangle/k_{\rm B}T_{\rm c}$ on pressure. In both In and Sn samples, $\alpha$ decreases with increasing pressure. In the indium sample, $\alpha$ changes from 1.89(1) at ambient pressure to 1.78(1) at $p\simeq 3.0$~GPa [Fig.~\ref{fig:alpha_vs_p}~(a)], while in the tin samples it decreases from 1.83(1) to 1.77(1) as pressure increases from 0 to $\simeq 2.9$~GPa [Fig.~\ref{fig:alpha_vs_p}~(b)]. Following the definition of $\alpha=\Delta/k_{\rm B}T_{\rm c}$ as the coupling strength (see above), this implies that pressure lowers the coupling strength and moves both In and Sn superconductors from the `intermediate-coupling' to the `weak-coupling' regime. Note that the effect of decreased coupling is also seen in $D(T/T_{\rm c})$ data [Figs.~\ref{fig:Bc_Deviation}~(e) and (f)] implying that pressure increases the deviation of $B_{\rm c}(T)$ curves from parabolic behavior.

The linear fits of $\alpha(p)$ presented in Fig.~\ref{fig:alpha_vs_p} yield $\alpha(p)=1.829(5)-0.037(2)\cdot p$ for In and $\alpha(p)=1.833(5)-0.024(4)\cdot p$ for Sn superconducting samples, respectively. Comparison with the universal weak-coupling BCS value $\alpha_{\rm BCS} = 1.764$, which sets the lower limit for the possible coupling strengths in phonon-mediated superconductors with an isotropic energy gap, implies that $\alpha(p)$ data would approach the $\alpha_{\rm BCS}$ value at $p \simeq 3.55$~GPa for In and $p \simeq 2.89$~GPa for Sn, respectively. If we were able to reach pressures higher than these limiting values, it might be possible to test the hypothesis that $\alpha$ values smaller than $\alpha_{\rm BCS}$ cannot be achieved. It is interesting to note that the last two points measured at pressures $p = 2.45$ and 2.87~GPa in the tin sample result in similar $\alpha = 1.765(10)$ values, which may indicate the saturation of $\alpha$ precisely at the weak-coupling BCS value $\alpha_{\rm BCS} = 1.764$. However, more measurements at higher pressures are needed to confirm or refute this observation.

Leavens and Carbotte, Ref.~\onlinecite{Leavens_CanJPhys_1972}, have shown that the effect of pressure on $\alpha$ in conventional phonon-mediated superconductors consist of two components. The first contribution arises from phonon hardening effects, while the second is determined by pressure-induced changes in the gap anisotropy. In Ref.~\onlinecite{Supplementary_Information} we describe a simple model that allows us to determine the pressure dependence of the phonon contribution to the coupling strength. The resulting phonon hardening and anisotropic contributions are presented in Fig.~\ref{fig:alpha_vs_p} in light blue and pink colors, respectively. Clearly, only part of the pressure effect on $\langle\Delta\rangle/k_{\rm B}T_{\rm c}$ can be attributed to the phonon contribution. Nearly 40\% of the effect is likely caused by the pressure-induced increase in the anisotropy of the superconducting energy gap. It is worth noting that the enhancement of gap anisotropy due to applied pressure was recently reported for elemental aluminum.\cite{Khasanov_Aluminum_PRB_2021}


To conclude, the pressure dependence of the thermodynamic critical field $B_{\rm c}$ in elemental superconductors indium (In) and tin (Sn) was studied using the muon-spin rotation/relaxation technique. With the pressure increase from 0.0 to $\simeq 3.0$~GPa, the coupling strength $\alpha=\langle\Delta\rangle/k_{\rm B}T_{\rm c}$ decreases: from 1.82 to 1.78 for indium and from 1.86 to 1.77 for tin, respectively. Linear fits  suggest that the coupling strength approaches the limiting weak-coupling BCS value $\alpha_{\rm BCS}=1.764$ at $p\simeq3.55$~GPa for In and $\simeq 2.89$~GPa for Sn. This implies that pressure lowers the coupling strength and moves both In and Sn superconductors from the `intermediate-coupling' to the `weak-coupling' regime. The analysis of $\alpha(p)$ data within the framework of the Eliashberg theory reveals that only part of the pressure effect might be attributed to the hardening of the phonon spectra, which is reflected in a decrease of the electron-phonon coupling constant. Nearly 40\% of the effect is caused by increased anisotropy of the superconducting energy gap.

\vspace{0.5cm}

Experiments were performed at the Swiss muon source (S$\mu$S, PSI Villigen, Switzerland) by using the GPD (General Purpose Decay) $\mu$SR spectrometer.


\begin{thebibliography}{99}


\bibitem{Lorenz_2005} B. Lorenz and C. Chu,  {\it High Pressure Effects on Superconductivity}, In: Narlikar, A.V. (eds) Frontiers in Superconducting Materials. Springer, Berlin, Heidelberg. (2005). \\
    \url{https://doi.org/10.1007/3-540-27294-1_12}

\bibitem{Schilling_book_2007} J.S. Schilling, {\it High-Pressure Effects}, In: Schrieffer, J.R., Brooks, J.S. (eds) Handbook of High-Temperature Superconductivity. Springer, New York, NY. (2007). \\
    \url{https://doi.org/10.1007/978-0-387-68734-6_11}

\bibitem{Schilling_JPCS_2008} J.S. Schilling and J. J. Hamlin, {\it Recent studies in superconductivity at extreme pressures}, J. Phys.: Conf. Ser. {\bf 121}, 052006 (2008).\\
    \url{https://doi.org/10.1088/1742-6596/121/5/052006}

\bibitem{Sun_Nature_2023} Hualei Sun, Mengwu Huo, Xunwu Hu, Jingyuan Li, Zengjia Liu, Yifeng Han, Lingyun Tang, Zhongquan Mao, Pengtao Yang, Bosen Wang, Jinguang Cheng, Dao-Xin Yao, Guang-Ming Zhang, and Meng Wang, {\it Signatures of superconductivity near 80 K in a nickelate under high pressure}, Nature {\bf 621}, 493 (2023).\\
    \url{https://doi.org/10.1038/s41586-023-06408-7}


\bibitem{Somayazulu_PRL_2019} Maddury Somayazulu, Muhtar Ahart, Ajay K. Mishra, Zachary M. Geballe, Maria Baldini, Yue Meng, Viktor V. Struzhkin, and Russell J. Hemley, {\it Evidence for Superconductivity above 260 K in Lanthanum Superhydride at Megabar Pressures}, Phys. Rev. Lett. {\bf 122}, 027001 (2019). \\
    \url{https://doi.org/10.1103/PhysRevLett.122.027001}

\bibitem{Drozdov_Nature_2019} A. P. Drozdov, P. P. Kong, V. S. Minkov, S. P. Besedin, M. A. Kuzovnikov, S. Mozaffari, L. Balicas, F. Balakirev, D. Graf, V. B. Prakapenka, E. Greenberg, D. A. Knyazev, M. Tkacz, and M. I. Eremets, {\it Superconductivity at 250 K in lanthanum hydride under high pressures}, Nature. {\bf 569} 528 (2019).\\
    \url{https://doi.org/10.1038/s41586-019-1201-8}

\bibitem{Semenok_arxiv_2024} Dmitrii V. Semenok, Ivan A. Troyan, Di Zhou, Andrei V. Sadakov, Kirill S. Pervakov, Oleg A. Sobolevskiy, Anna G. Ivanova, Michele Galasso, Frederico Gil Alabarse, Wuhao Chen, Chuanying Xi, Toni Helm, Sven Luther, Vladimir M. Pudalov, and Viktor V. Struzhkin, {\it Ternary superhydrides under pressure of Anderson's theorem: Near-record superconductivity in (La,Sc)H$_{12}$}, arXiv:2408.07477. \\
    \url{https://doi.org/10.48550/arXiv.2408.07477}


\bibitem{Franck_PRL_1968} J. P. Franck and W. J. Keeler, {\it Pressure Dependence of the Energy Gap of Superconducting Pb}, Phys. Rev. Lett. 20, 379 (1968).\\
    \url{https://doi.org/10.1103/PhysRevLett.20.379}

\bibitem{Svistunov_SovUspekhi_1987} V. M. Svistunov, M. A. Belogolovskii, and O. I. Chernyak, {\it Tunnel investigations of metals at high pressures}, Sov. Phys. Usp. {\bf 30}, 1 (1987).\\
    \url{https://doi.org/10.1070/PU1987v030n01ABEH002791}

\bibitem{Zhu_APL_2015} J. Zhu, Z.-X. Yang, X.-Y. Hou, T. Guan, Q.-T. Zhang, Y.-Q. Li, X.-F. Han, J. Zhang, C.-H. Li, L. Shan, G.-F. Chen, and C. Ren, {\it Tunneling spectroscopy of Al/AlO$_x$/Pb subjected to hydrostatic pressure}, Appl. Phys. Lett. {\bf 106}, 202601 (2015).\\
    \url{https://doi.org/10.1063/1.4921276}


\bibitem{Du_PRL_2024} F. Du, F. F. Balakirev, V. S. Minkov, G. A. Smith, B. Maiorov, P. P. Kong, A. P. Drozdov, and M. I. Eremets, {\it Tunneling Spectroscopy at Megabar Pressures: Determination of the Superconducting Gap in Sulfur}, Phys. Rev. Lett. {\bf 133}, 036002 (2024). \\
    \url{https://doi.org/10.1103/PhysRevLett.133.036002}


\bibitem{BCS_PR_1957} J. Bardeen, L. N. Cooper, and J. R. Schrieffer, {\it Microscopic Theory of Superconductivity},  Phys. Rev. {\bf 106}, 162 (1957).\\
    \url{https://doi.org/10.1103/PhysRev.106.162}

\bibitem{Tinkham_book_1975} M. Tinkham, {\it Introduction to Superconductivity} (Krieger Publishing company, Malabar, Florida, 1975).

\bibitem{Galkin_PSS_1969} A. A. Galkin, V. M. Svistunov, A. P. Dikii, {\it Effect of High Pressure on the Energy Gap of Indium and Thallium Superconducting Films}, Phys. Status Solidi {\bf 35}, 421 (1969). \\
    \url{https://doi.org/10.1002/pssb.19690350143}


\bibitem{Nedellec_SSComm_1973} P. N\'{e}dellec and R.J. Noer, {\it Tunneling in superconducting indium under pressure}, Solid St. Comm. {\bf 13}, 89 (1973). \\
    \url{https://doi.org/10.1016/0038-1098(73)90074-4}


\bibitem{Khasanov_Aluminum_PRB_2021} Rustem Khasanov and Igor I. Mazin, {\it Anomalous gap ratio in anisotropic superconductors: Aluminum under pressure}, Phys. Rev. B 103, L060502 (2021). \\
    \url{https://doi.org/10.1103/PhysRevB.103.L060502}

\bibitem{Leavens_CanJPhys_1972} C. R. Leavens and J. P. Carbotte, {Pressure Dependence of the Energy Gap Anisotropy in Aluminum}, Can. J. Phys. {\bf 50}, 2568 (1972).\\
    \url{https://doi.org/10.1139/p72-34}

\bibitem{Padamsee_JLTP_1973} H. Padamsee, J. E. Neighbor, and C. A. Shiffman, {\it Quasiparticle phenomenology for thermodynamics of strong-coupling superconductors},  J. Low Temp. Phys. {\bf 12}, 387 (1973).\\
    \url{https://doi.org/10.1007/BF00654872}

\bibitem{Johnston_SST_2013} D.C. Johnston, {\it Elaboration of the $\alpha-$model derived from the BCS theory of superconductivity} Supercond. Sci. Technol. {\bf 26}, 115011 (2013).\\
    \url{https://doi.org/10.1088/0953-2048/26/11/115011}


\bibitem{Khasanov_Three-wall-cell_HPR_2022} Rustem Khasanov, Ross Urquhart, Matthias Elender, Konstantin Kamenev, {\it Three-wall piston-cylinder type pressure cell for muon-spin rotation/relaxation experiments}, High Pressure Research {\bf 42}, 29 (2022). \\\url{https://doi.org/10.1080/08957959.2021.2013835}


\bibitem{Khasanov_HPR_2016} R. Khasanov, Z. Guguchia, A. Maisuradze, D. Andreica, M. Elender, A. Raselli, Z. Shermadini, T. Goko, F. Knecht, E. Morenzoni, and A. Amato, {\it High pressure research using muons at the Paul Scherrer Institute}, High Pressure Res. {\bf 36}, 140 (2016).\\
    \url{https://doi.org/10.1080/08957959.2016.1173690}

\bibitem{Khasanov_JAP_2022} Rustem Khasanov, {\it Perspective on muon-spin rotation/relaxation under hydrostatic pressure}, J. Appl. Phys. {\bf 132}, 190903 (2022). \\
    \url{https://doi.org/10.1063/5.0119840}

\bibitem{Khasanov_Bi-II_PRB_2019} R. Khasanov, M. M. Radonji\'{c}, H. Luetkens, E. Morenzoni, G. Simutis, S. Sch\"{o}necker, W. H. Appelt, A. \"{O}stlin, L. Chioncel, and A. Amato, {\it Superconducting nature of the Bi-II phase of elemental bismuth}, Phys. Rev. B {\bf 99}, 174506 (2019).\\
    \url{https://doi.org/10.1103/PhysRevB.99.174506}

\bibitem{Karl_PRB_2019} R. Karl, F. Burri, A. Amato, M. Doneg\`{a}, S. Gvasaliya, H. Luetkens, E. Morenzoni, and R. Khasanov, {\it Muon spin rotation study of type-I superconductivity: Elemental $\beta-$Sn}, Phys. Rev. B {\bf 99}, 184515 (2019).\\
    \url{https://doi.org/10.1103/PhysRevB.99.184515}

\bibitem{Khasanov_Ga-II_PRB_2020} R. Khasanov, H. Luetkens, A. Amato, and E. Morenzoni, {\it Structural phases of elemental Ga: Universal relations in conventional superconductors}, Phys. Rev. B {\bf 101}, 054504 (2020).\\
    \url{https://doi.org/10.1103/PhysRevB.101.054504}

\bibitem{Khasanov_AuBe_PRR_2020} R. Khasanov, R. Gupta, D. Das, A. Amon, A. Leithe-Jasper, and E. Svanidze, {\it Multiple-gap response of type-I noncentrosymmetric BeAu superconductor}, Phys. Rev. Research {\bf 2}, 023142 (2020).\\
    \url{https://doi.org/10.1103/PhysRevResearch.2.023142}

\bibitem{Poole_Book_2014} C. Poole, H. Farach, R. Creswick, and  R. Prozorov, {\it Superconductivity 3rd Edition} (Elseiver: Amsterdam, 2014).\\
    \url{https://doi.org/10.1016/C2012-0-07073-1}

\bibitem{deGennes_Book_1966} P.G. de Gennes, {\it Superconductivity of Metals and Alloys} (Benjamin, New-York, 1966).

\bibitem{Kittel_Book_1996} C. Kittel, {\it Introduction to Solid State Physics, 7th Ed.}, (Wiley, India, Pvt. Limited, 2007).

\bibitem{Prozorov_PRL_2007} R. Prozorov, {\it Equilibrium Topology of the Intermediate State in Type-I Superconductors of Different Shapes}, Phys. Rev. Lett. {\bf 98}, 257001 (2007).\\
    \url{https://doi.org/10.1103/PhysRevLett.98.257001}

\bibitem{Prozorov_NatPhys_2008} R. Prozorov, A. F. Fidler, J. R. Hoberg, and P. C. Canfield, {\it Suprafroth in type-I superconductors}, Nature Phys. {\bf 4}, 327 (2008).\\
    \url{https://doi.org/10.1038/nphys888}

\bibitem{Prozorov_PRAppl_2018} R. Prozorov and V. G. Kogan, {\it Effective Demagnetizing Factors of Diamagnetic Samples of Various Shapes}, Phys. Rev. Applied {\bf 10}, 014030 (2018).\\
    \url{https://doi.org/10.1103/PhysRevApplied.10.014030}

\bibitem{Supplementary_Information} The Supplementary Information part describes the TF-$\mu$SR data analysis procedure, the description of the phenomenoligical $\alpha-$model in describing $B_{\rm c}(T)$ data, as well the calculations of the 'phono hardening` contribution in the coupling strength $\alpha=\langle\Delta\rangle / k_{\rm B} T_{\rm C}$.

\bibitem{Egorov_PRB_2001} V. S. Egorov, G. Solt, C. Baines, D. Herlach, and U. Zimmermann, {\it Superconducting intermediate state of white tin studied by muon-spin-rotation spectroscopy}, Phys. Rev. B {\bf 64}, 024524 (2001).\\
    \url{https://doi.org/10.1103/PhysRevB.64.024524}

\bibitem{Leng_PRB_2019} H. Leng, J.-C. Orain, A. Amato, Y. K. Huang, and A. de Visser, {\it Type-I superconductivity in the Dirac semimetal PdTe2 probed by $\mu$SR}, Phys. Rev. B {\bf 100}, 224501 (2019).\\
    \url{https://doi.org/10.1103/PhysRevB.100.224501}

\bibitem{Gladisch_HypInteract_1979} M. Gladisch, D. Herlach, H. Metz, H. Orth, G. zu Putlitz, A. Seeger, H. Teichler, W. Wahl, and W. Wigand, {\it Muon spin rotation in superconductors}, Hyperfine Interact. {\bf 6}, 109 (1979).\\
    \url{https://doi.org/10.1007/BF01028778}


\bibitem{Grebinnik_JETP_1980} V. G. Grebinnik, I. I. Gurevich, V. A. Zhukov, A. I. Klimov, L. A. Levina, V. N. Maiorov, A. P. Manych, E. V. Mel'nikov, B. A. Nikol'skii, A. V. Pirogov, A. N. Ponomarev, V. S. Roganov, V. I. Selivanov, and V. A. Suetin, {\it Investigation of superconductors by the muon technique}, Zh. Eksp. Teor. Fiz. {\bf 79}, 518 (1980); [Sov. Phys. JETP {\bf 52}, 261 (1980)].

\bibitem{Beare_PRB_2019} J. Beare, M. Nugent, M. N. Wilson, Y. Cai, T. J. S. Munsie, A. Amon, A. Leithe-Jasper, Z. Gong, S. L. Guo, Z. Guguchia, Y. Grin, Y. J. Uemura, E. Svanidze, and G. M. Luke, {\it $\mu$SR and magnetometry study of the type-I superconductor BeAu}, Phys. Rev. B {\bf 99}, 134510 (2019).\\
    \url{https://doi.org/10.1103/PhysRevB.99.134510}

\bibitem{Maisurafze_PRB_2011} A. Maisuradze, A. Shengelaya, A. Amato, E. Pomjakushina, and H. Keller, {\it  Muon spin rotation investigation of the pressure effect on the magnetic penetration depth in YBa$_2$Cu$_3$O$_x$}, Phys. Rev. B {\bf 84}, 184523 (2011). \\
    \url{https://doi.org/10.1103/PhysRevB.84.184523}

\bibitem{Kozhevnikov_JSNM_2020} V. Kozhevnikov, A. Suter, T. Prokscha, and C. Van Haesendonck, {\it Experimental Study of the Magnetic Field Distribution and Shape of Domains Near the Surface of a Type-I Superconductor in the Intermediate State},  J. Supercond. Nov. Magn. {\bf 33}, 3361 (2020).\\
    \url{https://doi.org/10.1007/s10948-020-05576-1}

\bibitem{Carbotte_RMP_1990} J. P. Carbotte, {\it Properties of boson-exchange superconductors}, Rev. Mod. Phys. {\bf 62}, 1027 (1990). \\
    \url{https://doi.org/10.1103/RevModPhys.62.1027}

\bibitem{Finnemore_PR_1965} D. K. Finnemore and D. E. Mapother, {\it Superconducting Properties of Tin, Indium, and Mercury below 1$^{o}$~K}, Phys. Rev. {\bf 140}, A507 (1965).\\
    \url{https://doi.org/10.1103/PhysRev.140.A507}


\end{thebibliography}
\end{document}